# Effects of high pressure on the optical absorption spectrum of scintillating PbWO$_4$ crystals

D. Errandonea[a], D. Martinez-Garcia, R. Lacomba-Perales, J. Ruiz-Fuertes, and A. Segura

Dpto. de Fisica Aplicada-ICMUV, Universitat de Valencia, Edificio de Investigacion, c/Dr. Moliner 50, E-46100 Burjassot (Valencia), Spain

**Abstract**

The pressure behavior of the absorption edge of PbWO$_4$ was studied up to 15.3 GPa. It red-shifts at -71 meV/GPa below 6.1 GPa, but at 6.3 GPa the band-gap collapses from 3.5 eV to 2.75 eV. From 6.3 GPa to 11.1 GPa, the absorption edge moves with a pressure coefficient of -98 meV/GPa, undergoing additional changes at 12.2 GPa. The results are discussed in terms of the electronic structure of PbWO$_4$ which attribute the behavior of the band-gap to changes in the local atomic structure. The changes observed at 6.3 GPa and 12.2 GPa are attributed to phase transitions.

**PACs Numbers:** 71.20.Nr, 78.40.Fy, 62.50.+p

---

[a] Corresponding author: email: daniel.errandonea@uv.es, Tel.: (34) 354 4475, Fax: (34) 354 (3146)

Lead tungstate (PbWO$_4$) with scheelite structure is a wide-gap semiconductor with a band-gap energy ($E_g$) close to 4 eV [1]. It has attracted increasing interest in recent years because of its importance in applications as a laser host material [2], as a scintillator in high-energy physics detectors [3] and as an oxide ion conductor [4]. From the fundamental standpoint, PbWO$_4$ is also an interesting compound since it has two metastable crystal structures at ambient conditions: the tetragonal scheelite-type (stolzite) [5] and the monoclinic raspite-type [6]. In addition, a third phase (PbWO$_4$-III) can be quenched from high-pressure and high-temperature conditions [7]. High-pressure Raman [8 -11], x-ray diffraction [12 – 18], and neutron scattering [18] studies have been performed in stolzite and isostructural compounds. These studies allowed the determination of their high-pressure structural sequence. However, no information currently exists on how the electronic band structure of PbWO$_4$ is affected by the structural changes. In order to address this question, we have studied the pressure dependence on the absorption edge of PbWO$_4$ up to 15.3 GPa. The results will be presented and discussed here.

Five different samples were used for our optical measurements. They were cleaved along the {101} plane from a PbWO$_4$ single crystal grown by the Czochralski method starting from raw powders with 5N purity. An x-ray diffraction examination of the PbWO$_4$ crystal at ambient conditions showed that its diffraction pattern was in agreement with that of stolzite. The refined unit-cell parameters were $a$ = 5.459(7) A and $c$ = 12.042(9) A [5]. The thickness of the studied specimens ranged from 20 μm to 40 μm.

For optical absorption measurements in the UV-VIS-NIR range under pressure, the samples were loaded in a diamond-anvil cell with silicone oil or methanol-ethanol-water (16:3:1) as a pressure transmitting medium. The pressure was determined by the ruby fluorescence technique. The optical set-up consisted of a deuterium lamp, fused silica lenses, reflecting optics objectives and an UV-VIS spectrometer, which allowed for transmission measurements up to the absorption edge of IIA diamonds ($\simeq 5.5$ eV) [19].

The optical absorption spectra of $PbWO_4$, obtained at several pressures from the studied samples are shown in Fig. 1. The spectra measured at low pressures resemble those reported in literature at ambient conditions [1]. At 1 bar, a steep absorption starts at 3.9 eV. An absorption tail is also clearly seen between 3.3 eV and 3.9 eV. This tail is typical of $PbWO_4$. Its nature has been the subject of considerable debate [20] and is beyond the scope of this paper. This tail overlaps partially with the fundamental absorption which makes it hard to conclude whether the smallest band-gap is direct or indirect. It should be also mentioned that the observed absorption edge may also be excitonic in nature [21]. With this limitation in mind and in order to determine $E_g$, we shall follow Itoh considerations [1] and assume that the band-gap of scheelite $PbWO_4$ is of the direct-type and that the fundamental absorption edge obeys the Urbach's rule. In Fig. 1, it can be seen that the band-gap of $PbWO_4$ decreases with pressure. The pressure dependence obtained for $E_g$ is given in Fig. 2. There it can be seen that up to 6.1 GPa the reduction of $E_g$ appears to be almost linear, closing at a rate of –71(2) meV/GPa. A similar behavior has been observed for the scheelite-structured $PbMoO_4$ [8].

At 6.3 GPa we found changes in the color of the samples from colorless to light yellow. This color change is due to a collapse of the band-gap energy from 3.5 eV to 2.75 eV. A tetragonal-to-monoclinic phase transition has been reported to take place in $PbWO_4$ around 7 - 9 GPa [11, 17]. This transition induces changes on the tungsten-oxygen tetrahedron, which are expected to affect the band structure of $PbWO_4$ (see discussion below). Therefore, the model used to calculate the band-gap in the scheelite phase could not be applied beyond 6.3 GPa since the nature of the absorption edge in the high-pressure phase is currently unknown. We estimate the band-gap above 6.3 GPa from the tangents to the rising absorption edge and their intersections with the background. From 6.3 GPa to 11.1 GPa the absorption edge also moves to lower energies upon compression, with a pressure coefficient of –98(3) meV/GPa. At 12.2 GPa, a change of the shape and the pressure evolution of the absorption edge is observed. At the same time, the sample becomes orange in color. We attribute these changes to the occurrence of a second phase transition which has been observed in x-ray diffraction and absorption experiments near 15 GPa [17] and in Raman measurements at around 12.6 GPa [11, 22]. Beyond 12.2 GPa the absorption edge also red-shifts but with a pressure coefficient of –26(2) meV/GPa. In the scheelite-structured $BaMoO_4$, two transitions also take place upon compression [23]. The evolution of the absorption edge with pressure in the second high-pressure phase of $BaMoO_4$ has been measured [24], being found that it red-shifts at -21 meV/GPa, a value very close to the one observed by us after the second phase transformation in $PbWO_4$.

Based upon present knowledge of the electronic structure of scheelite $PbWO_4$, a qualitative approach towards the understanding of the presented results is suggested in the following. The scheelite structure is characterized by the tetragonal space group (SG) $I4_1/a$ [5], where the Pb and W sites have $S_4$ point symmetry. In this structure, the Pb cations are eightfold coordinated by O atoms, forming bisdisphenoids, and the W cations are fourfold coordinated by O atoms, forming nearly regular tetrahedra. With regards to the electronic structure, the valence and conduction bands near the band-gap are dominated by molecular orbitals associated with the $WO_4^{-2}$ ions [25]. In particular, the upper part of the valence band consists mainly of $O^{2-}$ $2p$ states, and the conduction band is dominated by $W^{6+}$ $5d$ states [25]. On the other hand, the $Pb^{2+}$ $6s$ states form a narrow band 1 eV below the bottom of the valence band [25]. A schematic diagram of the band structure of PbWO$_4$ is given in Fig. 3. According to the Ligand-Field theory [26], the splitting of the O$^{2-}$ 2$p$ states is dominated by the Coulomb attraction to the nearest W$^{6+}$ ions. The energy difference between the states with σ and π symmetries ($E_{p\sigma} - E_{p\pi}$) is proportional to $R_{WO}^{-3}$, where $R_{WO}$ is the W-O bond length. On the other hand, the splitting of the 5$d$ states of the W$^{6+}$ ion is dominated by the Coulomb repulsion from the four tetrahedral nearest neighbors' O$^{2-}$ ions. Where the energy difference between the states with $t_2$ and $e$ symmetry ($E_{t_2} - E_e$) is proportional to $R_{WO}^{-5}$. On increasing the pressure from 1 bar to 6.1 GPa, the W-O bond length is reduced by 2% [17], thus increasing the crystal field acting on 5$d$ and 2$p$ states. A consequence of this is a reduction in $E_g$, since the O$^{2-}$ 2$p(\pi)$ states shifts towards high energies faster than the W$^{6+}$ 5$d(e)$ states, causing a reduction of the energy difference between the 5$d(e)$ states at the bottom of the conduction band and the 2$p(\pi)$ states at the top of the valence band. Using the parameters given in Ref. [25] and the pressure dependence of $R_{WO}$ reported in Ref. [17], the decrease of the band-gap may be estimated. We found that, from 1 bar to 6.1 GPa $E_{p\sigma} - E_{p\pi}$ increases from 5 eV to 5.45 eV and $E_{t_2} - E_e$ increases from 1 eV to 1.15 eV. Therefore, at as a first approximation, $E_g$ is predicted to close at a rate of -50 meV/GPa. This value is similar to the value observed in the present experiments.

A transition from the tetragonal scheelite structure of PbWO$_4$ to the monoclinic fergusonite structure (SG I2/$a$) takes place at 6.9 GPa according to Raman studies [11, 22] and at 9 GPa according to x-ray diffraction and absorption measurements [17]. This transition involves a lowering of the point-group symmetry from 4/m to 2/m. It also occurs together with a distortion of the WO$_4$ tetrahedra [11, 17]. Upon further compression, the fergusonite structure of PbWO$_4$ is greatly distorted, changing the W-O coordination from 4, immediately after the phase transition, to 4+2, when the pressure exceeds the transition pressure by approximately 1 GPa [11, 17]. These changes of the crystalline structure are directly reflected in the electronic structure of PbWO$_4$, producing the collapse of $E_g$ that we observed at 6.3 GPa. In particular, the increase of the W-O coordination is coherent with the abrupt decrease of $E_g$. In this sense, it is well known that at 1 bar the photoluminescence of scheelite tungstates, with fourfold W coordination, is in the blue region while that of the wolframite tungstates, with 4+2 W coordination, is in the green region [27, 28]. The extrapolation of our data measured from 6.3 GPa to 11.1 GPa yields a band-gap energy of 3.38 eV for the high-pressure polymorph at 1 bar. On the other hand, according to recent *ab initio* calculations [29], the band-gap energy of raspite PbWO$_4$, in which each W is coordinated by six O, is estimated to be 0.7 eV smaller than that of scheelite PbWO$_4$ (4 eV). What is more, all the members of the double-perovskite structured tungstate family (e.g. Sr$_2$CaWO$_6$), which have a purely octahedrally coordinated W ion, have $E_g \simeq 3.5$ eV [30]. This fact supports our view that the collapse of the band-gap observed at 6.3 GPa is related to the W-O coordination changes produced after the previously reported pressure-driven phase transition [11, 17].

In order to close the discussion we would like to comment on the changes observed in the absorption edge at 12.2 GPa. Raman measurements locate the completion of a second pressure-induced phase transition in PbWO$_4$ at around 12.6 GPa [11, 22]. X-ray diffraction and absorption measurements locate this second transition at 15 GPa [17]. The transition occurs from the distorted monoclinic fergusonite structure to the more compact monoclinic PbWO$_4$-III structure (SG P2$_1$/*n*). A similar phase transitions is reported in BaMoO$_4$ [24]. We believe that the changes in the absorption spectrum observed at 12.2 GPa are also a consequence of an structural modification. In this case, they are probably caused by the second phase transition. The small decrease of $E_g$ during this transition is related to the small changes of W-O bond distances and angles in the $WO_6$ octahedra between the ferguson ite and $PbWO_4$-III phases since the ferguson ite phase acts in $PbWO_4$ as a bridge structure between fourfold and sixfold W coordination [11, 17]. Furthermore, the smaller decrease of $E_g$ in the $PbWO_4$-III phase when compared to the ferguson ite phase correlates well with the smaller decrease of the volume in the $PbWO_4$-III phase due to its larger bulk modulus [17]. The fact that the absorption spectra of the second-high pressure phase of $BaMoO_4$ [24] resemble those that we measured above 12.2 GPa in $PbWO_4$ and the fact that $E_g$ closes at a similar rate after the second transition in both compounds give support to our hypothesis. In addition to these considerations, we would like to mention here that the scheelite phase was reported to coexist with the high-pressure phase up to 10-11 GPa in a recent study [18]. Therefore, another possible interpretation for our present results is that the change observed in the band-gap at 12.2 GPa was an indication of the completion of the first transition. Further experimental and theoretical studies are being conducted to clarify this point.

In summary, absorption spectra of $PbWO_4$ were measured as a function of pressure up to 15.3 GPa. We observed a red-shift of the absorption edge under compression up to 6.1 GPa and abrupt changes of the absorption spectrum were found to take place at 6.3 GPa and 12.2 GPa. These changes were attributed to the occurrence of previously observed phase-transitions. The results are explained in terms of the crystal chemistry and electronic structure of $PbWO_4$. However, a definitive interpretation could only arise from high-pressure electronic structure calculations for the different structures of $PbWO_4$.

**Acknowledgements**: The authors thank Dr. P. Lecoq (CERN) for providing PbWO$_4$ crystals. This study was supported by the Spanish government MCYT (Grant No: MAT2004-05867 C03-01). D. Errandonea acknowledges the financial support from the MCYT of Spain and the Universitat of Valencia through the "Ramon y Cajal program. This manuscript benefited immensely from detailed reviews of F. J. Manjon and S. Gilliland.

[22] According to Raman studies the fergusonite and PbWO$_4$-III phases coexist from 7.9 GPa to 12.6 GPa. However, the peaks assigned to PbWO$_4$-III are very much weaker than those of fergusonite, suggesting than this is the dominant phase. Above 12.6 GPa, the transition to PbWO$_4$-III is completed and only the Raman peaks of it are observed.

[23] V. Panchal, N. Garg, and S. M. Sharma, J. Phys.: Condens. Matter **18**, 3917 (2006).

[24] D. Christofilos, J. Arvantidis, E. Kampasakali, K. Papagelis, S. Ves, and G. A. Kouroukliş *Proc. Joint 20th AIRAPT – 43th EHPRG Conference*, ed. by E. Dinjus and N. Dahmen, (Forschungszentrum Karlsruhe, 2005) T10-P99, ISBN 3-923704-49 6.

[25] Y. Zhang, N. A. W. Holzwarth, and R. T. Williams, Phys. Rev. B **57**, 12738 (1998).

[26] C. J. Ballhausen, *Introduction to Ligand-Field Theory* (McGraw-Hill, NY, 1962).

[27] M. Nikl, P. Bohacek, E. Mihokova, *et al.*, *Proc. 5th Intern. Conf. on Inorganic Scintillators and their Applications*, Moscow (Russia), p. 429 (2000).

[28] M. Itoh, and M. Fujita, Phys. Rev. B **62**, 12825 (2000).

[29] Y. Sun, Q. Zhang, T. Liu, and Z. Yi, phys. stat. sol. (b) **243**, 1248 (2006).

[30] H. W. Eng, P. W. Barnes, B. M. Bauer, and P. M. Woodward, J. Solid State Chem. **175**, 94 (2003).

**Figure captions**

**Figure 1:** Absorption spectra of scheelite PbWO$_4$ single crystals for different pressures.

**Figure 2:** Pressure dependence of the absorption edge of PbWO$_4$. Different symbols correspond to different samples. Lines are the least squares regression lines for the data.

**Figure 3:** Schematic diagram of the band structure of scheelite PbWO$_4$ using the notation of Ref. [26]. The shaded boxes are included to indicate that the discrete states of WO$_4^{-2}$ ions are broadened (creating bands) due to the crystal-field splitting and hybridization of the molecular orbitals in the bulk material. The position of the Pb 6*s* states is also indicated.

**Figure 1**

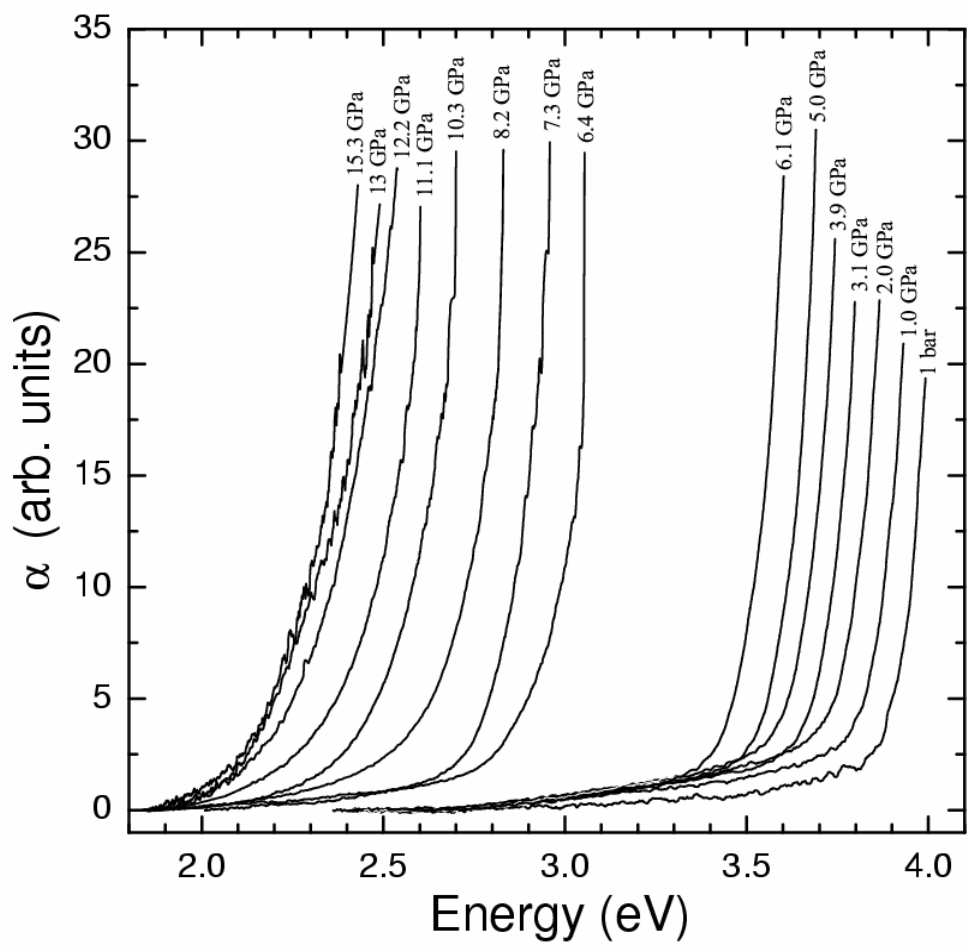

**Figure 2**

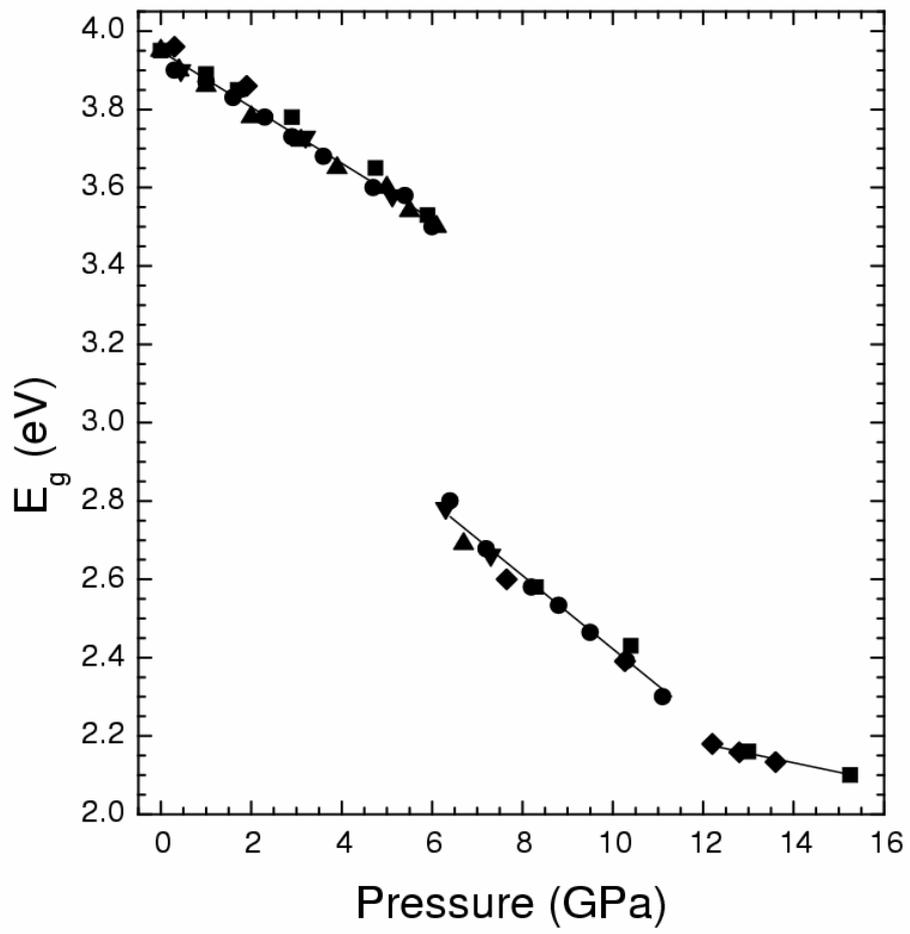

**Figure 3**

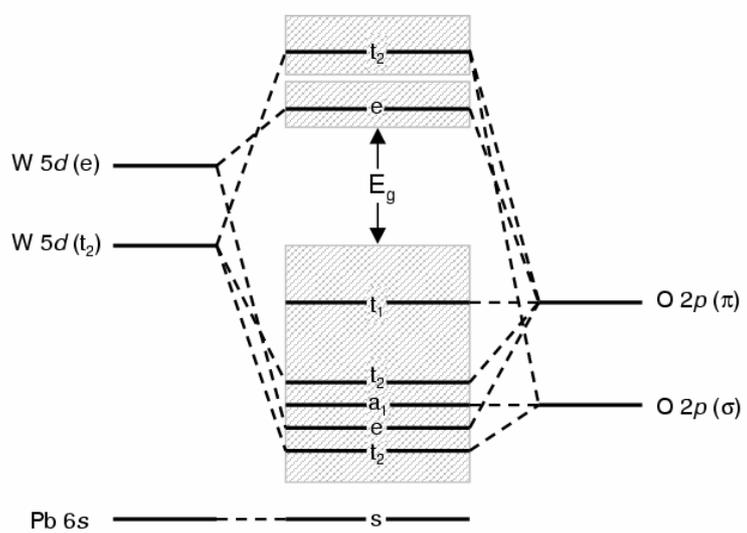